\documentclass[a4paper,12pt]{article}

\usepackage[dvipsnames]{xcolor}
\usepackage{amsmath}
\usepackage{amssymb}
\usepackage{dsfont}
\usepackage[english]{babel}
\usepackage[utf8]{inputenc}
\usepackage{times}
\usepackage{graphics}
\usepackage{graphicx}
\usepackage[T1]{fontenc}
\usepackage[official,right]{eurosym}
\usepackage{url}
\usepackage{eso-pic}
\usepackage{tikz}
\usetikzlibrary {arrows.meta,bending,positioning}
\usetikzlibrary{shapes.geometric, arrows}
\usepackage{eurosym}
\usepackage{enumerate,rotating,multirow,xcolor,booktabs,placeins}
\usepackage{algorithm}
\usepackage{algpseudocode}
\usepackage{appendix}
\usepackage{fancyhdr}
\usepackage{natbib}
\usepackage{xcolor}
\usepackage{colortbl}

\newcommand{\doi}[1]{\url{https://doi.org/#1}}

\def\qed{\hbox to\hsize{\hfill\vrule height 1.6ex width 1.5ex depth -.1ex}}

\usepackage{enumitem}
\newlist{steps}{enumerate}{1}
\setlist[steps, 1]{label = Step \arabic*:}

\title{Meaningful, Useful and Legitimate \\ Information in the Use of Index Numbers \\ for Decision Making}
\author{Fred Roberts$^\dag$, Helen Roberts$\ddag$, Alexis Tsouki\`as$*$ \\ \small{$^\dag$ DIMACS, Rutgers University} \\ \small{$\ddag$ Department of Mathematics, Montclair State University} \\ \small{$*$ LAMSADE-CNRS, PSL, Université Paris Dauphine}}
\date{}

\definecolor{maroon}{RGB}{128,0,0}
\begin{document}

%
%
%
%
%
%
%

\thispagestyle{empty}

\enlargethispage*{8cm}
 \vspace*{-38mm}

\AddToShipoutPictureBG*{\includegraphics[width=\paperwidth,height=\paperheight]{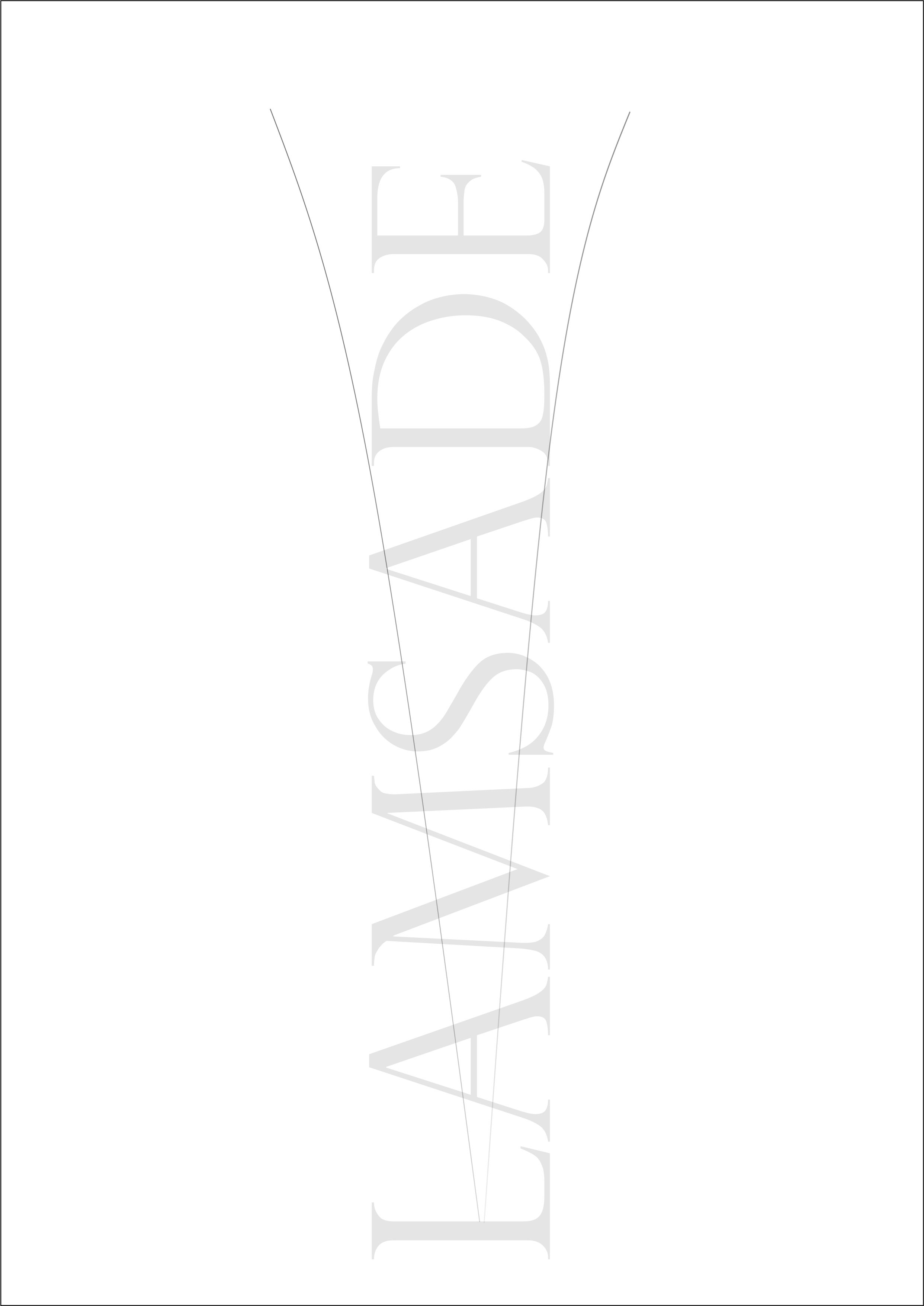}}

\begin{minipage}{24cm}
 \hspace*{-28mm}
\begin{picture}(500,700)\thicklines
 \put(60,670){\makebox(0,0){\scalebox{0.7}{\includegraphics{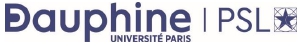}}}}
 \put(60,70){\makebox(0,0){\scalebox{0.3}{\includegraphics{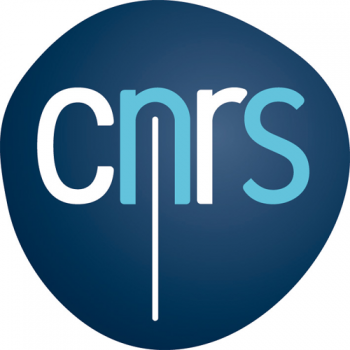}}}}
 \put(320,350){\makebox(0,0){\Huge{CAHIER DU \textcolor{BurntOrange}{LAMSADE}}}}
 \put(140,10){\textcolor{BurntOrange}{\line(0,1){680}}}
 \put(190,330){\line(1,0){263}}
 \put(320,310){\makebox(0,0){\Huge{\emph{411}}}}
 \put(320,290){\makebox(0,0){January 2025}}
 \put(320,230){\makebox(0,0){\Large{Meaningful, Useful and Legitimate Information}}}
 \put(320,210){\makebox(0,0){\Large{ in the Use of Index Numbers for Decision Making}}}
 \put(320,100){\makebox(0,0){\Large{Fred Roberts, Helen Roberts, Alexis Tsoukiàs}}}
 \put(320,670){\makebox(0,0){\Large{\emph{Laboratoire d'Analyse et Mod\'elisation}}}}
 \put(320,650){\makebox(0,0){\Large{\emph{de Syst\`emes pour l'Aide \`a la D\'ecision}}}}
 \put(320,630){\makebox(0,0){\Large{\emph{UMR 7243}}}}
\end{picture}
\end{minipage}

\newpage

\addtocounter{page}{-1}

\maketitle

\abstract{
Often information relevant to a decision is summarized in an index number. This paper explores conditions under which conclusions using index numbers are relevant to the decision that needs to be made.  Specifically it explores the idea that a statement using scales of measurement is meaningful in the sense that its truth or falsity does not depend on an arbitrary choice of parameters; the concept that a conclusion using index numbers is useful for the specific decision that needs to be made; and the notion that such a conclusion is legitimate in the sense that it is collected and used in a way that satisfies cultural, historical, organizational and legal constraints. While meaningfulness is a precisely defined concept, usefulness and legitimacy are not, and the paper explores properties of these concepts that lay the groundwork for making them more precise. Many examples involving two well-known and widely-used index numbers, body mass indices and air pollution indices, are used to explore the properties of and interrelationships among meaningfulness, usefulness, and legitimacy.}


\newpage

\section{Introduction}\label{intro}

Decision processes involving complicated issues within a society are driven by information retrieved, refined,
synthesized in order to be helpful for the ultimate issues to be decided. Often the relevant information is summarized in an index number that is used to help make a decision concerning such important topics as public health or air pollution.  One can ask: Under what conditions can we trust and use such index numbers when we need to make a decision? How does the answer depend on the kind of decision we need to make, the use we will make of the decision, or the societal context in which the decision will be made? Consider for example the use of index numbers to establish or assess policy initiatives. Under what conditions should one recommend to medical providers that a high ``body mass index'' should lead to medical intervention or that a high value on an air quality index should lead to shutting down some outdoor activities or some sources of pollution? For what kinds of decisions is the application of such index numbers useful? And would such index numbers be considered legitimate by some countries or societies and not others? These are the kinds of questions we will explore in this paper.

These types of questions
have been considered within the decision analysis community for a long time. See for instance \cite{LandryPascotBriolat83,LandryMalouinOral83,MeinardTsoukiasEJOR2018}. However, in recent times they have again emerged because of the massive diffusion of ``automatic'' or ``automatized'' devices, platforms or software that deliver critical information to decision makers in many different contexts such as bank loans, employment decisions, insurance coverage, salary increases, college admissions, predictive justice, etc. We are not going to survey this mass of literature \citep[the reader can have a look in][]{Tsoukias2021,Kiratetal2023}, but wish to emphasize that most of the discussion today revolves around the concepts of ``fairness'' or ``explicability'' (accountability) of what such devices recommend to their users. However, these concepts are only part of the more general issue of whether information produced for a certain decision process is ``appropriate,'' ``usable,''  ``acceptable,'' or ``relevant.'' We will use the term ``relevant'' to cover all of these attributes of information.

This paper is a first result of an ongoing research project between the DIMACS Center at Rutgers University and the LAMSADE at CNRS, PSL-Université Paris-Dauphine aiming at addressing the
kinds of questions raised above.

In this context, we present three concepts that underlie the relevance of information about index numbers: meaningfulness, usefulness, and legitimacy. One of these concepts, meaningfulness, has a precise mathematical definition and the others have a variety of relevant but primarily methodological approaches based on social and political sciences as well as statistical approaches. A key concept is \emph{meaningfulness} \citep[see][]{roberts79}: The truth or falsity of a statement using scales of measurement should remain unchanged if admissible (allowable) transformations of scale (such as from Kilograms to Pounds or degrees Centigrade to degrees Fahrenheit) are made.  Information that fails this condition is meaningless and can be misleading since it depends on a somewhat arbitrary choice of parameters such as unit or zero point. (As we will observe in Section 2.2, this definition of meaningfulness needs to be modified in special cases where not every acceptable scale can be obtained from another acceptable scale by an admissible transformation. The definition we are adopting will suffice for our discussion.)

Although meaningfulness of conclusions using information, and specifically index numbers, is necessary for the conclusions to be relevant in decision making, this is not a sufficient requirement. Information needs to
be \emph{useful}: It needs to
satisfy the demands for which it has been requested, of the person or organization that is expected to use it for some decision purpose. This introduces a subjective dimension that is not formally captured by meaningfulness, and that needs to complement the one of meaningfulness. However, once again although these two requirements of meaningfulness and usefulness are necessary, they are not sufficient. Information also needs to be \emph{legitimate}: The way it is collected, 
refined, and used needs to satisfy cultural, historical, organizational, legal constraints that condition the whole decision process. In contrast to the concept of meaningfulness, which
we have precisely defined, at least for the purposes of this article, the concepts of usefulness and legitimacy are not given specific definitions here, though there is relevant literature that will help us understand these concepts. In this paper, we explore some of the properties of the concepts of usefulness and legitimacy that will, hopefully, lay the groundwork to make them more precise in the spirit of the theory of measurement as explicated in the books by \cite{KrantzLuceSuppesTversky71}, \cite{LuceKrantzSuppesTversky90}, \cite{Suppesetal89}, and \cite{roberts79} and to better understand how they are related to meaningfulness.

To summarize, we will study the relevance to a decision of three distinct features we believe need to be satisfied:

\begin{enumerate}
  \item {\bf{Meaningfulness:} }Numerical information needs to represent consistently the empirical observations we perform.
  \item {\bf{Usefulness:}} Numerical information needs to be useful for the agent and the process for which has been constructed.
  \item {\bf{Legitimacy:}} Numerical information needs to take into account the social impact it may produce if and when used.
\end{enumerate}

We will
explore concepts of ``meaningfulness,''
``usefulness,''
and ``legitimacy''
 reflecting the above three features. It turns out that the second and third concepts, although intuitively easy to grasp, are far from being easy to formalize and, to make things more complicated, they might not be independent of each other or independent of meaningfulness. In order to understand better how such concepts can and should be used we explore in depth various conclusions/statements using two well-known and widely-used indices from different disciplines, Body Mass Indices and Air Pollution Indices, and consider whether these conclusions/statements are meaningful, useful, or legitimate. Our analysis will be multi-disciplinary, calling upon issues of public policy, social and behavioral science, medicine, public health, and environmental science, among others.
While some of our inquiry is exploratory, aiming at raising a number of key questions, we do give such specific results as determining which combinations of these three concepts are feasible and the meaningfulness, usefulness, and legitimacy of various operations on index numbers.

The paper is organized as follows. In Section 2 we discuss relevant literature, in Section 3 we analyze and discuss the BMI, while in Section 4 we analyze and discuss Air Pollution Indexes. Section 5 summarizes some key observations from our analyses,
providing some important conclusions from our analyses and questions the analyses have raised. Section 6 concludes, identifying some research directions that the project has left unanswered.

\section{Relevant Literature}\label{lit}

\subsection{Relevance}

As \cite{HuvilaEtAl19} point out, relevance has been a key concept in library and information science, but there is no consensus on what it means. \cite{StrassheimNasu} point out that concepts such as usefulness are central to the notion of relevance in library and information science. Huvila, et al. explore different kinds of relevance, user relevance and subject or topical relevance, something that arises in information retrieval \citep[see][]{Saracevic2016}. As Huvila, et al. note, some authors \citep[e.g.][]{MaoEtAl2016} have called topical relevance ``relevance'' and user relevance ``usefulness.'' Mao, et al. argue that relevance can be objectively measured, at least in information retrieval contexts, whereas usefulness is subjective. They also argue that relevance does not necessarily imply usefulness and ``user satisfaction.''

\subsection{Meaningfulness}

As already mentioned in the Introduction the notion of meaningfulness is relatively simple.
A statement involving a scale of measurement is considered meaningful if its truth or falsity is unchanged after an admissible transformation (one depending on the measurement scale adopted) is applied, More generally, a statement involving multiple scales of measurement is considered meaningful if its truth or falsity is unchanged after admissible transformations are applied (independently) to all scales in the statement.

The concept of meaningfulness 
was introduced in \cite{Suppes1959} and \cite{SuppesZinnes1963}
and was studied extensively in \cite{roberts79} \citep[see also the discussion in][and elsewhere]{Luce1978,Roberts1980}. 
Information that fails this condition is meaningless and can be misleading since it depends on a somewhat arbitrary choice of parameters such as unit or zero point. Thus, for example, a statement may be true in degrees Fahrenheit and false in degrees Centigrade: A temperature of 80 degrees Fahrenheit is twice a temperature of 40 degrees Fahrenheit, but using the corresponding Centigrade readings, a temperature of 26.67 degrees Centigrade is not twice a temperature of  4.44 degrees Centigrade. There is a long literature on mathematical analysis of meaningfulness and applications in a wide variety of fields \citep[see for example][]{Roberts1990,Roberts2014,Mahadevetal1998}.

As \cite{RobertsFranke1976} point out, the definition of meaningfulness needs to be modified if the scales of measurement arise from so-called irregular representations, ones in which not every acceptable scale can be obtained from another acceptable scale by an admissible transformation Semiorders provide an example.\footnote{Semiorders arise, for instance, when we are making judgments of preference but judgments of indifference are not transitive because indifference corresponds to closeness. Then we seek to find a utility function $f$ so that $x$ is preferred to $y$ ($xPy$) if and only if the utility assigned to $x$ is sufficiently greater than the utility assigned to $y$, i.e., for some constant threshold $\delta$, $xPy \iff f(x) > f(y) + \delta.$ See \cite{PirlotVincke97}, \cite{roberts79}}
The definition we have given is reasonably well accepted, at least to the extent that it is widely agreed that ``invariance'' is a desirable condition and that it is implied by ``meaningfulness,'' and it will suffice for our discussion. Not everyone likes the term meaningfulness. \cite{Mitchell1986}, for example, argues that the very term encourages us to disregard meaningless statements, when in fact there are occasions when we should not. For example, as we observe below, they might suggest errors in measurement. Others \citep[for example][]{FalmagneNarens1983} argue that the general definition is somewhat imprecise. There have been a variety of alternative formulations of the definition. Other definitions can be found, for example, in \cite{LuceKrantzSuppesTversky90} and \cite{Narens1981}.

Meaningfulness is strongly related to the notion of admissible transformation of measurement scales.
Following Stevens \citep[see][]{Stevens1946,Ste51,Stevens1959}, 
we will refer to different types of scales of measurement as defined by the class of admissible transformations. A scale of measurement is a \emph{ratio scale} if the admissible transformations are multiplication by a positive constant (e.g., mass, length) amounting to change of unit; \emph{interval scale} if the admissible transformations are multiplication by a positive constant and addition of a constant, the latter amounting to change of zero point (e.g., temperature); and an \emph{ordinal scale} if the admissible transformations are all monotone increasing functions (e.g., ranking of minerals by hardness). Thus, the admissible transformations $\phi$ for ratio scales are exactly all proportional transformations $\phi(f(x))=\alpha f(x)$ and for interval scales are exactly all affine transformations $\phi(f(x))=\alpha f(x)+\beta$, where $\alpha$  is a positive constant and $\beta$ is any constant. These are not the only scale types of interest, but the ones relevant to the examples we will give. 

From a measurement theory perspective it is meaningful to say that $x$ is $10\%$
 taller than $y$ since height (length) is measured on a ratio scale and so if $f(x) = 1.1\times f(y)$  is true, then this remains true if $f$ is replaced by $\alpha f$. However, if we measure the ``value'' of a person (for some attribute) or policy on an interval scale, to say that one person's or policy's value is 10\% higher than another's is meaningless, since the statement $f(x) = 1.1\times f(y)$ may be true with one choice of unit and zero point and false with another.
For example, if we measure the value of a climate change policy by the amount of temperature decrease it is predicted to achieve, then the statement that policy $x$ achieves a 10\% larger temperature decrease than policy $y$ may be true in Fahrenheit and false in Centigrade.

Note that it is meaningful to say that my
weight is 1000 times what it was a year ago, since the statement is certainly false no matter what units are used to measure
weight. Meaningfulness is different from truth. A meaningful statement that is so obviously false might be useless except that it might suggest an error in measurement or in recording the results. A false meaningful statement could also be useful in other ways, e.g., when it says that my body fatness is more than 10\% what it was a year ago. It could be false but useful in proving that I did not get that much fatter than I had been. Thus, the usefulness of a meaningful statement might depend on the use to which it was put.

The notion of meaningfulness extends to the use of measurement information for any type of statistic or any other way of computing
numbers for some decision support purpose. In general, it is meaningful to compare arithmetic means of measurements $f(x)$
 if $f$ defines a ratio scale, since the truth or falsity of the following statement is unchanged after multiplication of $f$ by a positive constant $\alpha$.

\[\frac{1}{n}\sum_{x\in A}f(x)>\frac{1}{m}\sum_{y\in B}f(y)\]

\noindent Similarly, one can see that it is even meaningful to compare arithmetic means if $f$ defines an interval scale since the truth or falsity of the comparison above remains unchanged if $f(x)$ is replaced by $\alpha f(x) + \beta$. However,
it is meaningless to compare arithmetic means if $f$ defines an ordinal scale. In this case, comparison of medians is meaningful.

\subsection{Measurement Invariance}

Invariance is the key concept underlying the theory of meaningfulness. The term \emph{``Measurement Invariance''} (\emph{MI}) is widely used in the literature, both as to measurement in the physical sciences and as to measurement in the social and political sciences, and has a variety of meanings. MI has aspects of meaningfulness, of usefulness, and of legitimacy, but it is not exactly any of these concepts. Typically, MI refers to results that make comparisons legitimate (not in our sense), e.g., between countries, states, groups, etc. \citep[][]{Gromadzki2019,LinEtAl2021}, from one time period to another \citep[][]{Najera2017,Mussotter2022}, or across two time points \citep[][]{NajeraEtAl2024}. Assessment of MI is heavily based in statistical methods. \cite{Najera2017} notes that \emph{``The statistical theory for assessing MI can be traced back to the late 1980s, although the computational capabilities to take MI examinations further is fairly recent,''} referring in particular to \cite{Meredith1993}. As \cite{NajeraEtAl2024} note, MI \emph{``is a statistical property concerned with the extent to which a given measurement model is equivalent across populations or time periods''} \citep[citing][]{MerredithTeresi2006,Meredith1993}. They use multiple group factor analysis (MGFA) to study invariance across survey rounds, and assess ``item reliability'' and ``criterion validity'' using ``confirmatory factor models'' and generalized binary models with a Poisson link, seeking ``statistical equivalence'' between results between two time periods. Meaningfulness is not statistical (though an analysis of degrees of meaningfulness might be of interest). Legitimacy and usefulness might be. \cite{Mussotter2022} also uses ``confirmatory factor analysis'' (CFA) to compare three models of patriotism and nationalism in Germany and to see the extent to which the models consistently fit the data. \cite{Gromadzki2019} explores the MI for studies of attitudes toward homosexuality across European countries, specifically deriving an approval measure, and also uses CFA to determine MI. He argues that \emph{``The validity of a measure in cross-country comparison studies requires satisfying two fundamental conditions. First of all, the construct should measure what it claims it measures. Second, the construct should be measured in the same way in each country.''} He uses Pearson correlation to measure the similarity between conclusions. \cite{LinEtAl2021}, using CFA, study measures of fear/anxiety about COVID-19 and seek to determine their MI across 11 different countries, specifically those speaking different languages. They also study MI across groups with different ethnicity, age, and gender. They argue that MI \emph{``is important for an instrument to help healthcare providers and researchers meaningfully compare an underlying concept''} such as a given COVID fear scale \emph{ ``between subgroups.''} Similar studies of MI of a COVID fear measure are in \cite{SawickiEtAl2022}, using 48 countries and  \cite{JovanovicEtAl2024} 
using 60 countries. \cite{MerredithTeresi2006} study a variety of levels of MI. For example, the weakest is metric or weak invariance, or regression slopes being invariant across groups. \emph{``Strong factorial invariance implies that the conditional expectation of the response, given the common and specific factors, is invariant across groups''} and \emph{``Strict factorial invariance implies that, in addition, the conditional variance of the response, given the common and specific factors, is invariant across groups.''} While not specifically about MI, another paper discussing invariance, with an emphasis on social science data, is by \cite{Uher22}. She argues that \emph{``physical technologies cannot be applied to the intangible research objects studied in psychology, social sciences and their applied fields. Instead, data about these study phenomena are often generated directly by persons (e.g., interviews assessments, observations), and pertinent measurement theories and quantification practices (e.g., psychometric theories, rating scales) have been developed largely independently from those of metrology,''} citing \cite{Mitchell2008}, \cite{Stevens1946}, \cite{Torgerson}.
Many examples in our paper
deal with such data, and so Uher`s arguments are particularly relevant. Uher argues that \emph{``Quantifications are meaningful only if the basic qualities to be studied for their possible divisible properties remain rather constant,''} and her use of the term \emph{``meaningful''} is closely related to the way we use the term. \cite{Uher22} argues that \emph{``Numerical data generated in psychology and social sciences often constitute differential summary scores that have no quantity meaning in themselves (e.g., construct indices). Instead, their meanings are created through between-case comparisons and are therefore always bound to the particular sample studied, thus precluding the numerical values' comparability across studies and disciplines.''} She asserts that \emph{``numerical traceability is key to making quantitative information such as index numbers useful, where numerical traceability requires documented connections of the assigned values to known quantitative standards''} and compares this to how \emph{``physical measurement units and scales are defined to establish an internationally shared understanding of physical quantities.''}

\subsection{Usefulness}

The paper by \cite{HuvilaEtAl19}  explores the concept of usefulness as used in the library and information science literature and states that \emph{``There is a need for better conceptual clarity in the literature regarding usefulness and related concepts.''} Huvila, et al. note that the most typical idea is that usefulness is \emph{``the degree to which something, often a system, enhances job performance or task completion.''} This suggests that there are degrees of usefulness, a topic we return to later. Along these lines, they point to relevant literature, e.g., \cite{Saracevic2016}.  Similarly, \cite{TsakPap2006} say that \emph{``usefulness is the degree to which a specific information item will serve the information needs of the user.''} Huvila, et al. point out that usefulness is different from \emph{usability}; a technological product, for example, can be very usable (``user friendly'') but not very useful, reflecting the degree to which the product will provide benefit to those using it for certain activities \citep[see also][]{VenkateshEtAl2012ConsumerAA}. The same item of information can be useful for some tasks and not for others. Huvila, et al. give the example of a cookbook, which might be useful for those wanting to prepare a delicious-tasting meal, but not for those who are seeking a healthy diet, or vice versa. Finally, Huvila et al. note that \emph{``usefulness is situational. It is not inherent to a specific piece of information, a particular system or service but is constructed in the interaction of a user and that what is being used,''} and they cite \cite{NoceraEtAl2007} for more on this topic.

\cite{CholvyPer19} study usefulness of responses in information retrieval, for example, which topics ``best match'' topics in a user query. They talk about a variety of dimensions related to usefulness in this context: \emph{``aboutness,''} which measures the \emph{``topical matching between a document and a user query''} \citep[][]{HuangSoergel2012}; \emph{``coverage,''} which measures \emph{``how strongly the user interests are included in a document;''} \emph{``appropriateness,''} which measures \emph{``how suitable a document is with respect to the user interests;''} \emph{``novelty,''} which measures \emph{``how novel is the document with respect to what the system has already proposed to the user.''} The paper provides technical definitions of these concepts. This discussion makes it clear that usefulness is a multidimensional concept and that it will be difficult to find a single index of usefulness.

\cite{HuangSoergel2012} speak about queries that will help a user get the most useful information \textemdash something akin to the concept of \emph{``value of information''} \emph{(VoI)}
widely studied in the literature that we discuss below. They discuss a binary approach that allows one to determine whether or not a given piece of information is useful or not, an ordinal approach that allows one to determine which of two pieces of information are more useful, and a numerical approach that gives a numerical measure of usefulness. They provide four postulates for usefulness and give an example of a usefulness measure that satisfies those postulates. The postulate/axiomatic approach is central to the theory of measurement that led to the concept of meaningfulness, and we can hope that ideas in our article will aid in developing this kind of approach to the concept of  usefulness.

\cite{LuGus1994} 
also relate the concept of usefulness to the well-studied topic of 
VoI.  There are different definitions of VoI and different technical approaches to calculating it. None of them are exactly relevant to what we have in mind by usefulness, since they depend heavily on concepts from statistical analysis, stochastic dominance, information theory, etc., and typically measure degrees of usefulness. \cite{BehBel23} describe VoI as a way to measure \emph{``the usefulness or worth of acquiring or possessing certain information''} and say that \emph{``It helps to make informed decisions by assessing the potential benefits that can be gained from obtaining specific information.''} VoI is useful in determining \emph{``whether it is worth investing time, effort, or money in order to gather additional information before making a decision. As such the VoI allows to weigh the potential benefits against the costs associated with obtaining or processing that information.''} They point out that the question of how to measure VoI in a single numerical value is \emph{``unresolved,''} but they introduce a method depending on tools from information theory.

\cite{JacksonEtAl2022} provide a summary of VoI methods with an emphasis on health care applications. They define VoI as \emph{``a decision-theoretic approach to estimating the expected benefits from collecting further information of different kinds in scientific problems based on combining one or more sources of data.''} They observe that researchers often seek to aid in understanding which of a variety of pieces of information contributes the most to uncertainty about a conclusion, and VoI is a tool to help determine which observations would be most valuable. Jackson, et al. describe VoI as a tool of Bayesian decision theory where a model using a parameter $\theta$ is used to make a decision, uncertainty about $\theta$ is described by a probability distribution, and we seek to minimize the loss (or maximize the benefit) of making the decision based on different values of $\theta$. Key concepts involved are expected value of perfect information (EVPI) (expected value of eliminating all uncertainty about $\theta$) and expected value of sample information (EVSI) (the value of seeking specific data in order to reduce uncertainty). This is a notion well developed in decision theory \citep[see][]{French88} where VoI is defined as the difference between the expected utility without new information and the expected utility with the new information (typically the guess of an ``oracle'' about the ``real'' state of nature). \cite{Tuffaha} notes that EVPI and EVSI are difficult to calculate, though he points out recent new approaches using regression-based methods,  importance sampling methods, Gaussian approximation methods and moment matching methods.  \cite{RungeEtAl23} calculate EVPI by using a method based on \emph{``the algebra of the expected value of perfect information''} and separately calculating the relevance of the uncertainty to a decision and the magnitude of uncertainty.

\cite{Tuffaha}, writing in the context of pharmacology, defines VoI as a tool that \emph{``quantifies the expected value of research in reducing decision uncertainty to inform whether a decision can be made based on existing evidence or if additional evidence is required and worthwhile.''} He points out that \emph{``the additional benefits from a new research study may not justify its costs in terms of the direct research costs (e.g. site and recruitment costs) and the opportunity costs from the benefits forgone when the adoption of a promising health technology is deferred while the additional information is being collected.''} Then, for example by calculating EVSI, we can determine which sample size helps us to choose research studies with the highest expected returns on research investments.

In the literature, there is increasing interest in the distinction among statistical significance, practical significance, and clinical significance, with the latter two concepts sometimes used interchangeably \cite[see][]{Carpenteretal2021}. \cite{kirk2or83} say that \emph{``practical significance is concerned with whether the result is useful in the real world''.} So, this literature may also be helpful in understanding usefulness. Kirkwood and Sarin discuss ways to measure effect, with higher effect presumably being related to greater usefulness. \cite{Peeters2016} also relates practical significance to effect sizes.

\subsection{Legitimacy}

The term ``legitimacy'' is often used in ways that are not reflective of the way we are using the term. For example it is used in the literature in discussions of ``legitimate and reliable'' sources, as in \cite{Woods21}: \emph{``A legitimate source is one that is what it purports to be, and a reliable source is one that can be trusted to have valid or accurate information.''} However, as noted, there is no
trusted way of determining if a source is legitimate and reliable, though peer-reviewed sources are the most likely to fit these criteria. The United Nations Refugee Agency UNHCR also talks about legitimate sources as reliable ones\footnote{\url{https://help.unhcr.org/slovakia/safe-online/information/} Accessed 30/11/2024.}, and talks about the challenge of finding ``legitimate information'' about active conflicts in this age where you cannot trust everything you find over the Internet. \cite{Boyd20} talks about legitimacy of data and says that this is not just about quality and accuracy, but more about the belief \emph{``that those data are sound, valid, and fit for use.''} This is somewhat related to what we have in mind when we use the term ``legitimacy,'' reflecting the fact that some data may not be fit for use because of the possibility that principles of society or culture may be violated. Boyd talks about how data can be messy and filled with a variety of errors for various reasons, e.g., improper calibration of instruments, bias, or being taken from one context into another. In the latter sense, we are talking about a lack of ``invariance,'' which is more in the spirit of meaningfulness than legitimacy. In a similar study, \cite{Cowden2024} study similarities and differences between survey items in a study of personal wellbeing, noting that research on wellbeing has been criticized for its emphasis on factors from \emph{``Western, educated, industrialized, rich, democratic''} contexts. They point out that \emph{``researchers have found the meaning of salient terms in the wellbeing literature (e.g., happiness) often varies across cultures and languages''}\textemdash essentially implying that conclusions might not be legitimate in our sense of the word. As they point out, \cite{SmithTW2004} argues \emph{``that comparing human wellbeing across countries should be sensitive to potential variation in how individuals in different contexts understand the survey items that are presented to them.''} \cite{Najera2017} studies indices of poverty in Mexico, and makes the point that \emph{``The underlying assumption is that poverty is measured equivalently across populations, i.e. the indicators utilized to construct a multidimensional index are invariant manifestations of poverty across the groups or countries of interest. \ldots Multidimensional poverty measures rely on series of indicators to produce an index. They are combined in some way to compute a synthetic estimate of poverty. However, these indicators should be comparable in order to result in a valid and comparable measure across countries or groups.''}` Of course, if different cultural or legal matters lead to the indices omitting certain relevant features, then this underlying assumption fails and the information in the index is not legitimate in the sense that we are using the term. Najera mentions the potential difference between countries with \emph{``different standards of living or preferences.''}

Business ethics provide a different kind of example of issues of legitimacy. As \cite{ScholtensDam2007} argue, \emph{``We find that there are significant differences among ethical policies of firms headquartered in different countries.''} They observe that \emph{``Business ethics, as part of culture, does not happen in vacuum or isolation. It takes place in a social and cultural environment that is being governed by a complex set of laws, rules and regulations, formal values and norms, codes of conduct, policies, and various organizations.''} Similar conclusions and rationales are given by \cite{VitollaEtAl2021}. The Investopedia Team\footnote{\url{https://www.investopedia.com/ask/answers/040715/how-do-business-ethics-differ-among-various-countries.asp}. Accessed 30/11/2024.} points out that \emph{``Business practices that would be illegal, or at least frowned upon, at home are often allowed or at least tolerated elsewhere''} and \emph{``Many developing nations have lax insider trading laws. In some Latin American countries, bribery and kickbacks are a regular part of doing business.''}  Thus, it is noted, business decisions about whether to give gifts, whether to dump waste, whether to allow children to work long hours, {though based on the same index numbers, might be legitimate in some countries and not in others.

\section{Obesity/Body Mass Index}\label{bmi}

Health indices play an important role in designing health guidance for patients and their medical providers and for setting public health policy. In this section, we discuss an index of obesity, the \emph{Body Mass Index (BMI),} that is widely used in guidance provided by government health agencies such as the U.S. Centers for Disease Control and Prevention (CDC). We discuss the meaningfulness, usefulness, and legitimacy of different observations using the BMI.

It has long been felt that body fatness or adiposity is an indicator of potential medical problems, including high blood pressure, high cholesterol, Type 2 diabetes, coronary heart disease, stroke, and some types of cancers. However, existing ways to measure body fatness (adiposity) can be expensive or require specially trained personnel, and methods are difficult to standardize \citep[][]{CDC2022}. Some of those methods are skinfold thickness measurements, underwater weight measurement, bioelectrical impedance, and dual-energy x-ray absorptiometry \citep[][]{CDC2022}. Thus, metrics like percentage of body fat might not be \emph{usable} under any of these various ways of obtaining them since they are expensive to obtain, difficult to measure, and difficult to standardize, and if they are not usable, then they are  \emph{useless}. Usability is a necessary condition for usefulness, but not a sufficient one. The difficulty to standardize might make these metrics fail a condition like Measurement Invariance and might also make the metrics \emph{illegitimate}, since one person doing the measurements in one setting or country might come up with a different index than another person doing the measurement in a different setting or country. {In general, failure of Measurement Invariance certainly suggests illegitimacy if the failure is due to different social, cultural, or regularity matters. However, if the ways to measure body fatness lead to a ratio scale, as seems likely with things like skinfold thickness, underwater weight measurement, etc., then it is \emph{meaningful} to make comparisons such as body fatness of one person is 10\% higher than body fatness of another. Meaningfulness depends on the scales used to describe the data, not on the procedure used to gather the data or the characteristics of the population the statement using metrics or indices is describing or directed at. However, because of difficulty of measurement, e.g., for skinfold thickness, meaningful statements such as $f(x) = 1.1\times f(y)$ might be true in one technician's measurements and false in another's, which might make this piece of information useless, depending on the level of accuracy required in an application.

The body mass index, BMI, was popularized by Ancel Keyes in the early 1970s, using an idea originally due to Quetelet in the 19th century \citep[see][]{Rasmussen2019}. BMI is defined as the ratio of weight W (in Kg) to the square of the height H (in meters):

\[BMI = \frac{W}{H^2}\]

\noindent A person with BMI 30.0 or higher is considered obese (obesity is defined as excess of adipose tissue - \cite{Wellensetal1996}); someone with BMI between 25.0 and 29.9 is considered overweight. In contrast to measures of fatness, this is easy to measure and standardize. In that sense, it is usable. To see if it is useful,
one must ask whether it relates to adiposity and whether it does what it was designed to do, which is to predict certain kinds of diseases such as cardiovascular disease. So, in that sense, its usefulness must be determined. In Section 2.4, we discussed the notion of value of information, and there discussed the idea that VoI weighs the potential benefits against the costs associated with obtaining or processing information. Since the cost of getting weight and height information is so small, and the potential benefit in terms of health evaluation relatively high, this suggests that using the BMI is useful - at least for certain applications. Usefulness includes several components: ease of use (usability),
appropriateness for intended use, some extent of Measurement Invariance, and value of information in terms of benefits over costs.

\subsection{Comparisons Using BMI}

Are comparisons of BMI meaningful? If $BMI(x)$ is the body mass index of individual $x$ and $BMI(x,t)$ is BMI of $x$ at time $t$, it is meaningful to say that

\begin{eqnarray}
  BMI(x) & > & BMI(y) \\
  BMI(x) & = & 2\times BMI(y) \\
  BMI(x,t) & = & 1.2\times BMI(x,t-1)
\end{eqnarray}

\noindent These statements follow because multiplication of weight $W$ by a positive constant and of height $H$ by a (possibly different) positive constant leaves the statement true if it was initially true and false if it was initially false. If they are true with weight in Kg and height in meters, they are also true with weight in Kg and height in centimeters, or with weight in grams and height in inches (though why we would want to use a mix of metric and non-metric units is in question). However, it is not meaningful to say that x is obese, i.e., that $BMI(x)\geq 30$, without specifying units used. Of course, units are usually understood to be the standard units if they are not mentioned.

In comparing indices after different kinds of experimental treatments, we often try to compare averages. It is meaningful to say that the average BMI of one group of $n$ individuals is greater than the average BMI of another group of $m$ individuals. That is the case because multiplication of $W$ and of $H$ by positive constants (independently chosen) leaves the truth or falsity of the following statement unchanged.

\[\frac{1}{n}\sum_{x\in A}BMI(x)>\frac{1}{m}\sum_{y\in B}BMI(y)\]

It is also meaningful to say that the average BMI of a group of individuals has gone down by 10\% after a certain treatment. The latter is the case because, as before, multiplication of $W$ and of $H$ by positive constants (independently chosen) leaves the truth or falsity of the following statement unchanged.

\[\frac{1}{n}\sum_{x\in A}BMI(x)=1.1\times\frac{1}{m}\sum_{y\in B}BMI(y)\]

Comparing averages can get tricky, as is well known and has been discussed in \cite{Roberts2012}. This point warrants repeating in our context. For example, to return to the different ways of measuring body fatness, we observed that skinfold thickness might be difficult to measure, leading to variations from one measurement to another. In this case we might want to average over several measurements or over several measurers. This is a possible way to get around situations  where Measurement Invariance fails. Suppose $SF_i(x)$ is the skinfold thickness of person $x$ obtained by the $i^{th}$ measurer. Then we might ask whether the following statement is meaningful:

\[\frac{1}{n}\sum^n_{i=1}SF_i(x)>\frac{1}{n}\sum^n_{i=1}SF_i(y)\]

\noindent (Note that here we average over the same number $n$ of measurers on both sides of the statement, in contrast to the situation where we are comparing average measurements in one group of $n$ elements to the average in a second group of $m$ elements.) If all the $SF_i$ are measured on a ratio scale, as here, then this statement is meaningful as multiplication of both sides by the same $\alpha$ leaves the truth or falsity unchanged. However, could it possibly be that the $SF_i$ have independent units? In this case we would see if truth or falsity is unchanged after multiplication of $SF_i$ by a positive constant $\alpha_i$ that could be different for each $i$. It is easy to find such $\alpha_i$ so that the truth or falsity changes after multiplication of $SF_i$ by $\alpha_i$, resulting in the comparison

\[\frac{1}{n}\sum^n_{i=1}\alpha_i SF_i(x)>\frac{1}{n}\sum^n_{i=1}\alpha_i SF_i(y)\]

\noindent This makes comparison of arithmetic means meaningless in this situation. But is it reasonable to allow the different measurers to use independent scales? It could happen if there is no standardization. Whether or not this makes sense in this application, the point is that comparison of averages may lead to meaningless statements. In some applications, it clearly could make sense. Consider the hypothetical case where we are measuring improved health of animals, and one measurer decides to measure the improved health in terms of weight gain and a second in terms of height gain.

The observation that arithmetic mean comparisons might be meaningless has led some to suggest using geometric means rather than arithmetic means. Then the comparison becomes:

\[\sqrt[n]{\prod^n_{i=1} SF_i(x)}>\sqrt[n]{\prod^n_{i=1}SF_i(y)}\]

\noindent This is meaningful even if each measurer's ratio scale unit can be changed independently, since this comparison is true if and only if the following comparison is true:

\[\sqrt[n]{\prod^n_{i=1}\alpha_i SF_i(x)}>\sqrt[n]{\prod^n_{i=1}\alpha_i SF_i(y)}\]

\noindent For a further discussion of this topic, including applications in air pollution measurement \citep[see][]{Roberts2012}. However, just because comparisons using geometric means are meaningful does not make them useful. One can question whether they are.

\subsection{Different Ways to Measure BMI}

Note that the same statements (1), (2), and (3) are meaningful if $BMI = W/H^2$  or if $BMI = W/H^3$ or even $W/H^{500}$ . This underscores the need to distinguish meaningfulness from usefulness. Note that if BMI were taken to be $W + H$, then none of (1), (2) and (3) would be meaningful, since they might be true with Kg and meters but false with Pounds and centimeters.

One can ask why $W/H^2$ is preferable to $W/H^3$. The latter is sometimes called the \emph{Ponderal index}. This comes down to a discussion of growth. As an organism doubles in length or height, the surface area increases fourfold, but its volume and mass increases by a factor of eight, it is $2^3$. Now the organism needs eight times the biologically active tissue to support the surface area of its respiratory organs. Moreover, it has eight times the mass to support on its legs, which are dependent on cross-sectional area that has only grown by four. The compensation is to think of appropriate growth compensating in proportion, with volume/mass only growing by a factor of four\footnote{See ``Allometry'', 
\url{https://en.wikipedia.org/w/index.php?title=Allometry&oldid=1181258397}. Accessed 30/10/2023.}, i.e., $2^2$. This kind of discussion belongs to the field of allometry and is beyond the scope of this article \citep[see][for details]{Smith1968,Taylor2010}.

\subsection{BMI for Different Populations}

If an athlete has a BMI of 30 or higher, he or she would be called obese. But that statement, while meaningful, may not be so useful for athletes because the BMI might be higher because of increased muscularity rather than increased body fatness \citep[][]{CDC2022}. Thus, for certain populations, BMI is less useful than for others. In other words,
there is
need for understanding \emph{degrees of usefulness}. As noted in Section 2.4, Value of Information is one approach to measuring degree of usefulness. However, as we have observed, the concept of degree of usefulness is not quite the same, as it depends heavily upon concepts of statistical analysis, stochastic domination, and information theory.

BMI is interpreted differently for children than for adults since they are growing. The formula used is the same, but because the amount of body fat changes with age, and because it differs between boys and girls, the guidance as to what defines overweight or obesity changes
 \citep[][]{CDC2022}. The value of BMI defining obesity now depends on a reference population of children of a given age and sex, with obesity defined as having a BMI at or above the $95^{th}$ percentile in this population. Thus, if $x$ is a boy of age 12 and $y$ is a boy of age 13, it is meaningful to say that $BMI(x)>BMI(y)$, but this is not a useful comparison since it is comparing apples to oranges. The statement is, of course, meaningful and also useful if $x$ and $y$ are of the same age. Similarly, if $x$ is a boy of age 12 and $y$ is a girl of age 12, this statement is both meaningful and not useful. In short, usefulness can depend on the population a statement refers to, and it is possible that a statement that is similar might be useful for some comparisons and not for others. On the other hand, to say that a boy of age 12 is obese is a meaningful statement since if $BMI(x)$ is at least as high as that of 95\% of other boys of age 12 in the reference population, this statement remains true or false if weight and height are multiplied by appropriate positive constants. The statement is also, presumably, useful, and the usefulness of saying that an individual is obese is not dependent on the population that the individual belongs to (though the truth or falsity of the statement does depend on the population). Another comment: The choice of a 95\% threshold is related to usefulness, not meaningfulness. A 75\% threshold could just as easily be used and the resulting class of obese boys of age 12 would be much larger. But now the conclusion of obesity might not be as useful since it might not trigger the need for behavioral or medical intervention. It is clearer that exceeding the 95\% threshold should be a trigger than that a 75\% threshold should. The degree of usefulness of some information is related to the probability that it will lead to action or a decision, which in turn is related to Value of Information.

Concluding that an individual is obese can lead to behavioral or medical intervention, e.g., diet, exercise, or medication. However, these medical-based decisions may disregard cultural factors. In some societies, at least historically, obesity was valued. \cite{Pollock1995} gives the example of Tahitians who valued obesity and fattened up people to make them more sexually attractive and ``lusty and high spirited.'' She also gives the example of the island of Nauru, where young women were fattened up because fat was associated with fertility and beauty. Thus, for some cultures, the conclusion that a person is obese would not lead to medical intervention, but instead might be valued. For these, the conclusion from use of the BMI to medically intervene would not be
legitimate.

Christian Scientists generally view disease and illness as a mental issue, not a physical one, and so resort to prayer rather than medicine to cure disease\footnote{Harvard Divinity School ``Religion and Public Life: Christian Scientists in the Courts''. \url{https://rpl.hds.harvard.edu/religion-context/case-studies/minority-america/christian-scientists-courts}. Accessed 30/10/2023.}. Some believe that prayer will help in weight loss\footnote{Holzworth, G. ``A Prayerful Perspective On Obesity''. Christian Science Journal. 28/08/2006. \url{https://sentinel.christianscience.com/shared/view/1pqj5wso5jg}. Accessed 30/10/2023.}. However, this does not make the conclusion that they are obese illegitimate. What it does is make the resulting recommendation to use medicine illegitimate, because it violates religious principles.

\subsection{Correlation between BMI and Other Metrics}

It is of interest to see if two indices are associated and/or correlated, and this can be measured by calculating a correlation coefficient. For example, it is of interest to see if BMI and \%BF (percent of body fat) are correlated, since \%BF is a measure of fatness. If they are highly correlated, then BMI could be used as a proxy for \%BF. There are a number of different correlation coefficients that are in use, and in particular in the study of BMI. One such is \emph{Kendall's} $\tau$, which depends only on order of values. It is the least sensitive of the methods to calculate correlation since it does not use any
actual calculated values. Consider a population $P$ of individuals ranked by BMI and also by \%BF. Disregard ties. Then for each pair of individuals $x$ and $y$, say the rankings are \emph{concordant} if both rankings have $x$ over $y$ or both have $y$ over $x$, and say they are \emph{discordant} otherwise. If $C$ is the number of concordant pairs and $D$ the number of discordant pairs, then\footnote{Glenn, S. ``Kendall's Tau (Kendall Rank Correlation Coefficient)'', From Elementary statistics for the rest of us! \url{https://www.statisticshowto.com/kendalls-tau/}. Accessed 30/10/2023.}

\[\tau=\frac{(C-D)}{(C+D)}\]

\noindent Since both BMI and \%BF are ratio scales, and certainly at least ordinal scales, $C$ and $D$ are unchanged after admissible transformation of scales, and so is $\tau$. Thus, to say that BMI and \%BF have a correlation of, say, 0.65, is a meaningful statement. But whether a correlation of 0.65 is high enough to be able to conclude that BMI is a \emph{useful} proxy for \%BF is a matter for discussion. The answer may depend on the discipline. For instance, some feel that in physics, a correlation must be at least 0.95 or at most -0.95 to be useful, in chemistry at least 0.9 or at most -0.9, in biology at least 0.7 or at most -0.7, and in the social sciences at least 0.6 or at most -0.6 \footnote{Jost, S., ``Linear correlation'', IT223 Course Documents, Depaul University, ND. \url{https://condor.depaul.edu/sjost/it223/documents/correlation.htm}.  Accessed 29/10/2023.\label{Jost}}. In medicine, it would seem to be more like physics. One of the challenges is to understand what correlations are high enough to make a correlation useful in different applications.

Another way to measure correlation is \emph{Spearman's rank correlation}\footnote{University of Newcastle. Numeracy, Mathematics and Statistics, Academic Skills Kit: Strength of Correlation,  \url{https://www.ncl.ac.uk/webtemplate/ask-assets/external/maths-resources/statistics/regression-and-correlation/strength-of-correlation.html}, Accessed 29/10/2023.\label{newcastle}} $\rho$. Here, you rank order the observations in each of two variables (e.g., BMI and \%BF) being compared for $n$ different data points (e.g., individuals). For the $i^{th}$ data point, you let $d_i$ be the difference between the two ranks. Then Spearman's rank correlation is given by

\[\rho=1-\frac{6\sum^n_{i=1}d_i^2}{n(n^2-1)}\]

\noindent If the data are on an ordinal, interval or ratio scale, then the corresponding ranks do not change when there are admissible transformations of scale. Hence, in this case, as with Kendall's $\tau$, it is meaningful to say that the correlation between BMI and \%BF is 0.65 or some other specific value. Usefulness would be judged in the same way as for Kendall's $\tau$.

Another way to measure correlation is \emph{Pearson's correlation}, which applies to parametric data where we compare two values assigned to individuals and which is used when both variables are normally distributed. Because the correlation calculation is affected by extreme values, it is recommended not to use this correlation when the data are not normally distributed.  However, in the case of BMI and \%BF, the data sets are usually so large that normality is a reasonable assumption. This is a recommendation that has to do with the appropriateness of a statistical test, but not about meaningfulness of a conclusion. If $x_1, x_2, \cdots, x_n$ are the values obtained from one measure for $n$ individuals in a population and $y_1, y_2, \cdots, y_n$ are the values obtained from a second measure, then the Pearson correlation $r$ is given by

\[r=\frac{\sum_i(x_i-\bar{x})(y_i-\bar{y})}{\sqrt{\sum_i(x_i-\bar{x})^2\sum_i(y_i-\bar{y})^2}}\]

\noindent where $\bar{x}$ is the average of the $x_i$ and $\bar{y}$ is the average of the $y_i$\footnote{See footnote \ref{newcastle}.}.
If the $x_i$ and $y_i$ are measured on a ratio scale, then $r$ remains unchanged if all $x_i$ are multiplied by the same positive $\alpha$ and all $y_i$ are multiplied by the same positive $\beta$. This means that it is meaningful to say that the correlation $r$ is a given value.

In \cite{Smalleyetal1990} the Pearson correlation between $BMI = W/H^2$ and \%BF over a given population was calculated as 0.84 for women and 0.70 for men. In \cite{Wellensetal1996} the Spearman rank correlation between BMI and \%BF was calculated to be 0.79 for women and 0.65 for men. Is it meaningful to say that the correlation for women is higher than for men? Yes, because correlation does not change with admissible transformations of scale. In both cases the correlation seems high (though not as high as expected, say, in physics), which perhaps suggests that BMI might comfortably be used instead of \%BF, a conclusion about
usefulness. However, might there be different thresholds for ``high'' for different populations, e.g., adults vs. children? There do not seem to be any obvious concerns about legitimacy.

\subsection{Sensitivity/Specificity}

The \emph{sensitivity} of a test gives the number of true positives (number correctly identified as positive) over the number of positives, and the \emph{specificity} gives the number of true negatives (number correctly identified as negative) over the number of negatives. In \cite{Smalleyetal1990} it was concluded that for men, the sensitivity of the BMI index, taking \%BF as accurate, is 44.3\%, which means that less than half of the men identified as obese in terms of \%BF were identified as obese by BMI. For men, the specificity of BMI taking \%BF as accurate was determined to be 90.1\%, which means that 90.1\% of those men not obese were correctly identified as not obese by BMI. (Thus, the sensitivity is not particularly Measurement Invariant, but the specificity is to a large extent.) These conclusions are certainly meaningful. The suggestion is that BMI is not very useful in identifying men who are actually obese, but quite useful in identifying men who are not obese. So, while it is meaningful to say that a man is obese or that a man is not obese, the usefulness of these two conclusions differs. It is another peculiarity of the concept of usefulness that a conclusion and its negation, while both meaningful, can differ in terms of usefulness. Similar conclusions about sensitivity and specificity were obtained in \cite{Wellensetal1996}, where sensitivity was calculated to be 44.0\% for men and specificity was calculated to be 93.0\% for men.

\subsection{Relationship between BMI and other Metrics: Regression}

In linear regression, we seek to estimate the relationship between two variables by fitting a straight line to data. For example, in \cite{Hannanetal1995} a linear regression of \%BF as a function of BMI for adolescent females was calculated as

\[\%BF = 1.446\times BMI - 3.6\]

\noindent They also reported how much of the variation in \%BF is explained by the fitted regression line, in other words, the proportion of the total variation in the dependent variable \%BF that can be attributed to the independent variable BMI. This proportion is called $r^2$. In fact, it is the square of the Pearson correlation coefficient. The higher the $r^2$, the better the regression line fits. They reported an $r^2$ of 0.576.

For adult females,
\cite{Hannanetal1995} reported

\[\%BF =1.816\times BMI - 13.99\]

\noindent and $r^2$ = 0.659. Since $r^2$ is higher in the second example, this suggests that BMI is a better substitute for \%BF for adult females than for adolescent females. This conclusion is meaningful in the sense that if change of scale takes place, the same regression line will result. However, whether it is useful depends on whether the linear regression line is in fact a good fit and on whether the $r^2$ obtained in either of the cases is high enough to suggest that BMI is a good substitute for \%BF. It may also be useless if the difference between the two values of $r^2$ is not statistically significant. How do we address that question?

A high $r^2$ suggests a large positive or large negative linear association. Still, what makes a large positive or negative linear association depends on uses to be made of the conclusion and on the discipline, as we have previously observed for correlations as well. For example,
in physics an $r^2$ of 0.9 is considered by some\footnote{See footnote \ref{Jost}.} necessary for a linear association to be ``meaningful''. However, this notion of meaningfulness corresponds more to our concept of usefulness (and to the concept of Value of Information, since you get less value for the information if correlation is low). For chemistry the $r^2$ necessary to be considered ``meaningful'' is 0.8, for biology 0.5, and for social sciences 0.35\footnote{Again, see footnote \ref{Jost}.} . One can take this one step further and note that the decisions one wants to make may require different levels of $r^2$ to make us comfortable. Deciding on a medical treatment may require an $r^2$ of 0.9, whereas deciding on what telephone to buy may only require an $r^2$ of 0.4. So, usefulness of a conclusion involving index numbers can depend on the decision the conclusion is used for.

Returning to the BMI/\%BF example, we note that even the higher $r^2$ obtained, that for the case of adult females, is likely not enough to conclude that BMI is a good substitute for \%BF in medical applications. Thus, at least based on that study, we may conclude that BMI is not useful for these medical purposes.

\subsection{BMI as a Predictor of Disease}

A variety of studies have investigated whether BMI is a predictor of different kinds of disease or different kinds of biomarkers that are thought to be predictive of disease. For instance, \cite{Willetetal2006} calculated Pearson correlations between BMI and biological markers that reflect adiposity, for example: FBG (fasting blood glucose); HDL (high-density lipoprotein cholesterol); SBP (systolic blood pressure); TG (triglyceride). Changes in the levels of these biomarkers are four of the most important physiological consequences of obesity. For men, the correlation of HDL with BMI was -0.32, in other words, suggesting that an increase in BMI was correlated with decrease in HDL (higher HDL is better). The correlation of FBG with BMI for men was 0.19. So, would you decide to try to lower your BMI because you had learned it had a correlation of -0.32 with HDL and you had heard that lower HDL was connected to a higher risk of developed coronary heart disease? Would you make a decision to try to lower your BMI because you learned it had a correlation of 0.19 with FBG, which in turn was predictive of diabetes? How useful these conclusions are will depend on a wide variety of assumptions about your behavior, priorities, etc. By way of contrast, would your physician feel comfortable in recommending lowering your BMI based on these results?
In both cases it would depend on the level of your BMI, but perhaps differently. The resulting decision depends not on the meaningfulness of the results, but on some judgments of usefulness of the results and, perhaps also, about how the results will be applied, e.g., to change your diet vs. to start you on a new medication. For example, the correlation combined with the level of your BMI might be sufficiently high to recommend a change of diet but not necessarily to recommend starting on a new medication.

In \cite{Willetetal2006} the bioelectrical impedance analysis (BIA) was studied as a measure of \%BF and BIA (and hence \%BF) was compared to BMI as predictors of these biomarkers FBG, HDL, SBP, and TG. The authors concluded that BMI is a better, cheaper, and easier measure to use and its correlations with these biomarkers were comparable to those of BIA. As we noted earlier, ways to measure degree of usefulness
are worth developing, and here is another example where BIA and BMI have similar correlations with \%BF, but using BMI would be much more usable and therefore potentially much more useful because of ease of calculation (not because of usefulness in terms of correlation with a metric; there are a variety of interpretations of usefulness).

\cite{Hunteretal1997} found that BMI was generally as good as or better at predicting HDL-C (amount of cholesterol found in HDL), serum TGs, and SBP compared with percent body fat measured by computed tomography. Based on these kinds of investigations, is there enough evidence to continue to
explore computed tomography for measuring \%BF as an alternative to BMI as a
predictor of these biomarkers? Again, this is not a question of meaningfulness. The answer may depend on whether the decision is to be made by an individual physician or a government agency, so again the usefulness may depend on the user. If there are government agencies involved and, perhaps, some regulations or guidelines, then legitimacy also becomes an issue.

A related analysis connecting BMI to cholesterol levels is discussed in \cite{Garrisonetal1980}. The authors concluded that different lipid levels are correlated with BMI differently for men vs. women and for people of different age groups. For example, for men ages 20-29, the
Pearson correlation with HDL is -0.304, whereas for women at these ages it is -0.171. For men ages 40-49, the correlation with HDL is -0.167. Both of these correlations would generally be considered weak negative correlations\footnote{see footnote \ref{newcastle}}, independent of the discipline of the study. Even though it is meaningful, would a physician feel comfortable telling a 45-year-old man with a high BMI that his
likelihood of having a lower HDL is less than it was when he was 25? Or not pushing a 25-year-old woman to lower her BMI as much as pushing a 25-year-old man with a similar BMI to do so? Again, these are questions addressed to the usefulness of a conclusion, and the answer may differ as to the significance of the decision involved. And it should also be noted that usefulness needs to be taken in context. For example, how such questions are addressed may depend on additional measures, such as abdominal circumference or muscle mass.

\cite{Denkeetal1993} used regression to show that increasing BMI is associated with detrimental lipoprotein profiles' higher total cholesterol and non-HDL cholesterol levels and a higher total-HDL cholesterol ratio, with higher triglyceride levels and lower HDL cholesterol levels. They argued that: \emph{``Since obesity is a modifiable risk factor, it deserves special attention in both the population approach and the high-risk strategy approach for coronary heart disease risk reduction through control of cholesterol levels''} and so there is argument for pushing people to reduce their BMI. However, they did not report $r$ or $r^2$.

Similar results have been obtained for cancer. For example, \cite{Bhaskaranetal2014} investigated BMI as a predictor of cancer risk and found it linearly associated with a number of major cancers including cancers of the uterus, gallbladder and kidney.

\section{Air Pollution Measurement}\label{aqi}

There are a variety of pollutants such as carbon monoxide ($CO$), hydrocarbons ($HC$), nitrogen dioxide ($NO_2$), sulfur dioxide ($SO_2$), ozone ($O_3$), and (various kinds of) particulate matter ($PM$). From the time that
air pollution measurement was in its infancy, there was a goal of finding a single pollution index based on the levels of emissions of different kinds of pollutants. This could help in comparing strategies for pollution control and take into account tradeoffs and it might also help in making recommendations for individual behavioral responses to different pollution conditions. The issues were and still are rather complex. For example, it might be possible that a policy that leads to the increase in some pollutants might lead to an overall decrease in pollution if we can measure using one index number. In this section, we discuss a variety of air pollution indices and explore the meaningfulness, usefulness, and legitimacy of conclusions using these indices, for example for developing or evaluating pollution reduction policies or for giving advice to individuals as to when to limit outdoor activities when the air is ``bad.''

\subsection{Comparisons Using Weight of Pollutants in Emissions}

A simple idea is to measure total weight of emissions of pollutant $i$ over a fixed period of time in a given volume of air and sum over $i$ to define a pollution index. The pollutant concentration is measured in units such as milligrams per cubic meter $mg/m^3$ or micrograms per cubic meter $\mu/m^3$ (which is usually used for particulate matter) or parts per million by volume (ppm) (which is usually used for $CO$ or $O_3$) or parts per billion by volume (ppb) (which is usually used for $NO_2$ and $SO_2$). There are conversions that take the metric in milligrams per cubic meter and change it to ppm or ppb by volume, and vice versa\footnote{``Air Pollutant Concentrations'',
\url{https://en.wikipedia.org/wiki/Air_pollutant_concentrations}. Accessed 09/11/2023.}.  Let $e(i,t,k) =$ total weight of emissions of pollutant $i$ over the $t^{th}$ time period and due to the $k^{th}$ source or measured in the $k^{th}$ location. Then let $A(t,k)$ be the sum over all $i$ of $e(i,t,k)$. Using the weight-based index $A$, \cite{Walther1972} concluded things like: \\
 - Transportation is the largest source of air pollution, with stationary fuel combustion (especially by electric power plants) second largest. \\
 - Transportation accounts for over 50\% of all air pollution. \\
 - CO accounts for over half of all emitted air pollution.

Conclusions such as these are meaningful if we use a standard measure of pollutant concentration such as milligrams per cubic meter or ppm or ppb by volume since volume defines a ratio scale.  For they can be written as:

\[A(t,k) > A(t,k')\]
\[A(t,k_r) > \sum_{k\neq k_r}A(t,k)\]
\[\sum_{t,k}e(i,t,k)> \sum_{t,k}\sum_{j\neq i}e(j,t,k)\]

\noindent But are they useful? A unit of mass of $CO$ is far less harmful than a unit of mass of $NO_2$. This suggests that simply summing as in $A(t,k)$ is not a useful measure of pollution. At one point early in the days of air pollution measurement, U.S. Environmental Protection Agency standards based on health effects for a 24-hour period allowed 7800 units of $CO$, 330 units of $NO_2$, 788 of $HC$, 266 of $SO_2$, 150 of $PM$  \citep[see][]{EPA1971}. These are \emph{Minimum acute toxicity effluent tolerance factors} (MATE criteria) or \emph{tolerance factors} \citep[see][]{BabcockNagda1973,Hangerbrauck1977}. The tolerance factor is the level at which adverse effects are known or thought to occur. There are other issues. For example, some of these pollutants are more serious in the presence of others, e.g., $SO_2$ are more harmful in the presence of $PM$ \citep[see][]{Napca1969}. Also, the products of chemical reactions of the different pollutants can be damaging. Oxidents such as ozone are produced by $HC$ and $NO_2$ reacting in the presence of sunlight. These measures disregard both of these complications, which suggests that using the pollution index A, for example to set standards for emissions or make other air pollution policy decisions, fails the usefulness criterion. Using a variety of sub-indices (as here) is common with index numbers. But conclusions from index numbers, though meaningful, can be useless if they disregard the kinds of interactions/interdependencies among the factors measured by the sub-indices as the ones here.  What about legitimacy? While the conclusions we have discussed are useless, they are legitimate; the way they are obtained does not seem to violate cultural, historical, organizational, or legal constraints.

\subsection{Toward an Index that Considers Health Impacts}

Let $\tau(i)$ be the tolerance factor for the $i^{th}$ pollutant. Let the \emph{severity factor} be $1/\tau(i)$. One idea is to weight the emission levels (in mass) by the severity factor and get a weighted sum.  This amounts to using the indices $1/\tau(i) \times e(i,t,k)$ and summing these to get $B(t,k)$, which is called \emph{Pindex} \citep[see][]{Babcock1970}:

\[B(t,k) = \sum_i\frac{1}{\tau(i)}e(i,t,k)\]

\noindent Under \emph{Pindex}, transportation was still the largest source of pollutants, but now accounting for less than 50\%. Stationary sources fell to fourth place. $CO$ dropped to the bottom of the list of pollutants, accounting for just over 2\% of the total \citep[see][]{Walther1972,BabcockNagda1973}. \emph{Pindex} was introduced in the San Francisco Bay Area in the 1960s when they first tried to seriously measure pollution \citep[see][]{BayAreaPollution1968,SauterChilton1970}. It is easy to see that these conclusions are again meaningful as long as all emission weights are measured in the same units. But, as before, we can ask if they are legitimate or useful.

\emph{Pindex} amounts to the following:  For a given pollutant, take the percentage of a given harmful level of emissions that is reached in a given period of time, and add up these percentages over all pollutants. (The sum can be greater than 100\% as a result.) If 100\% of the $CO$ tolerance level is reached, this is known to have some damaging effects.  But \emph{Pindex} implies that the effects are equally severe if levels of five major pollutants are relatively low, say 20\% of their known harmful levels. There is thus some doubt that this index of pollution gives useful results. One other comment: In the early days of air pollution measurement, reported severity factors differed from study to study. One reason was that air quality standards were not all laid out for the same time period; some for one hour, some for eight hours, etc. There were differing opinions as to how to extrapolate the standards to the same time period, e.g., 24 hours. Thus, using \emph{Pindex} again failed on the usefulness criterion, since the ways it was measured were inconsistent. Of course, this also shows that \emph{Pindex violated Measurement Invariance. And if Measurement Invariance fails, usefulness is questionable.} Note that, similarly to the case of the weight-based index $A$, using \emph{Pindex} does not necessarily seem illegitimate.

\subsection{Air Quality Index: AQI}\label{aqi1}

A method currently in use in the U.S. is the Air Quality Index (\emph{AQI}) (sometimes known as the pollutant standard index or PSI\footnote{AirNow. AQI. Basics. \url{https://www.airnow.gov/aqi/aqi-basics/}. Accessed 30/10/2023.}). The AQI has been issued by the U.S. Environmental Protection Agency since 1976 with regular updating to reflect information about health effects\footnote{AirNow. Using the AQI. \url{https://www.airnow.gov/aqi/aqi-basics/using-air-quality-index/}. Accessed 30/10/2023.\label{airnow}}. Variants of the AQI are now used around the world. We return to the variants below and observe that the differences may lead to some AQI-based conclusions being legitimate in some countries and not in others. In the U.S., the AQI aims to provide an easy-to-understand daily report on air quality in a format that is the same from state to state in the country. The AQI focuses on health effects that an individual might experience within a few hours or days after breathing polluted air \citep[see][]{PlaiaRuggieri2011}.

The AQI assigns a number between 1 and 500 for AQI sub-indices for each of five pollutants: $PM$, $CO$, $SO_2$, $NO_2$, and
$O_3$. The sub-indices are calculated by converting measured pollutant concentrations (e.g., in $\mu/m^3$ or $ppm$ or $ppb$) to a uniform index based on the health effects associated with a pollutant. Those health benchmarks are established by the Environmental Protection Agency and updated regularly\footnote{Minnesota Pollution Control Agency. Understanding the AQI.  \url{https://www.pca.state.mn.us/search?search=understanding+the+air+quality+index}. Accessed 30/10/2023.\label{minnesota}}. The overall AQI is reported as the highest of the AQI sub-indices. One day (or hour) it could be due to ozone, another to $CO$.

The level for each pollutant is reported in one of six categories of increasing seriousness as shown in Table \ref{aqitable}. In most parts of the U.S, forecasts are given for $PM$ and $O_3$, with some places also giving forecasts for $CO$ and $NO_2$ and $SO_2$\footnote{see footnotes \ref{airnow} and \ref{minnesota}}. It is certainly meaningful to say that the AQI for a given pollutant $i$ is in a more serious category today than it was yesterday.

But what if the AQI score for ozone of 209 was highest yesterday and the AQI for $CO$ of 230 was highest today? Is it meaningful to say that the overall air quality AQI was higher today (worse) than yesterday? The EPA has created AQI values for the different pollutants from known information about health effects. For example, we know that the moderate level of health effects for ozone ranges between 0.055 $ppm$ to 0.070 $ppm$. The moderate level of AQI for any pollutant ranges from 51 to 100, so 0.055 $ppm$ corresponds to an AQI value of 51 for ozone, and 0.070 $ppm$ corresponds to an AQI value of 100. For $ppm$ values between 0.055 and 0.070, the AQI values are obtained by interpolation between 51 and 100. The same kind of thing works for $CO$, where the limits of the moderate level of AQI are between 4.5 and 9.4 $ppm$. So, for $CO$, 4.5 $ppm$ corresponds to AQI of 51, 9.4 $ppm$ to AQI of 100, and $ppm$ values between 4.5 and 9.4 are obtained by interpolation. Similarly, for some kinds of $PM$s ($PM_{2.5}$)\footnote{Particulate matter is classified by size, specifically diameter, or particles. Those with diameter at most 10 microns define $PM_{10}$ and those with diameter at most 2.5 microns define $PM_{2.5}$. The former can be deposited on the surface of the lungs and the latter into the deeper parts of the lungs. While both result in health effects, $PM_{2.5}$ is responsible for the greater proportion of health effects due to air pollution. California Air Resources Board, Inhalable Particulate Matter and Health ($PM_{2.5}$ and $PM_{10}$). \url{https://ww2.arb.ca.gov/resources/inhalable-particulate-matter-and-health}. Accessed 08/01/2024.}, the moderate level of AQI ranges between 12.1 and 35.4 milligrams per cubic meter. In other words, scales are set up so that a given AQI for ozone is in the same ``place'' relative to health effects as the same AQI for $CO$ and for $PM$ \citep[see][]{EPA2018,PlaiaRuggieri2011}. Volume defines a ratio scale. Changing volume measurement from cubic meter to cubic feet or cubic meter to cubic kilometer would not result in changes of the mapping between concentration and AQI values, whether at the boundaries or in between via interpolation. So, the conclusion that the overall air quality AQI was
worse today than yesterday seems to be meaningful.

\begin{table}
 \small
  \centering
   \begin{tabular}{cccl} \hline
     Daily AQI & Levels of & Values & ~~~~~~~~~~~~~~Description of Air Quality \\
     Color & Concern & of Index & ~ \\ \hline \hline
     \rowcolor{green} Green & Good &	0 to 50	& Air quality is satisfactory, and air pollution \\
     \rowcolor{green} ~ & ~ & ~ & poses little or no risk \\ \hline
     \rowcolor{yellow} ~ & ~ & ~ & Air quality is acceptable. However, there may \\
     \rowcolor{yellow} Yellow &	Moderate & 51 to 100 & be a risk for some people, particularly those \\
     \rowcolor{yellow} ~ & ~ & ~ & who are unusually sensitive to air pollution. \\ \hline
     \rowcolor{orange} ~ & Unhealthy & 101 to & Members of sensitive groups may experience \\
     \rowcolor{orange} Orange &	Sensitive & 150 & health effects. The general public is less likely \\
     \rowcolor{orange} ~ & Groups & ~ & to be affected. \\ \hline
     \rowcolor{red} ~ & ~ & 151 to & Members of the general public may experience \\
     \rowcolor{red} Red	& Unhealthy	& 200 & health effects; members of sensitive groups \\
     \rowcolor{red} ~ & ~ & ~ & may experience more serious health effects. \\ \hline
     \rowcolor{purple} \textcolor{white}{Purple} & \textcolor{white}{Very} & \textcolor{white}{201 to} & \textcolor{white}{Health alert: The risk of health effects is} \\
     \rowcolor{purple} ~ & \textcolor{white}{Unhealthy} & \textcolor{white}{300} & \textcolor{white}{increased for everyone.} \\ \hline
     \rowcolor{maroon} \textcolor{white}{Maroon} &	\textcolor{white}{Hazardous} &	\textcolor{white}{301 and} & \textcolor{white}{Health warning of emergency conditions:} \\
     \rowcolor{maroon} ~ & ~ & \textcolor{white}{higher} & \textcolor{white}{everyone is more likely to be affected.} \\ \hline
   \end{tabular}
\caption{Source https://www.airnow.gov/aqi/aqi-basics/}\label{aqitable}
\end{table}

Is the AQI useful for decision making? Here, the U.S. Environmental Protection Agency (through AirNow.gov) offers guidance. For example, one can ask: If the AQI forecast for tomorrow is 120, should I go out to exercise tomorrow? Along with the AQI, recommendations for interpretation are issued\footnote{AirNow. Air Quality Guide for Ozone. \url{https://document.airnow.gov/air-quality-guide-for-ozone.pdf}. Accessed 30/10/2023.}. For example, for ozone, the recommendation is this for a score of 101-150, which is unhealthy for \emph{``Sensitive Groups,''} groups that \emph{``include people with lung disease such as asthma, older adults, children and teenagers, and people who are active outdoors.''} For such groups: \emph{``Make outdoor activities shorter and less intense. Take more breaks. Watch for symptoms such as coughing or shortness of breath. Plan outdoor activities in the morning when ozone is lower. People with asthma: Follow your asthma action plan and keep quick-relief medicine handy.'' }For \emph{``Everyone else'':}
\emph{``Consider making outdoor activities shorter and less intense.''} Similarly, for a level between 201-300, which is unhealthy for everyone, the recommendation is: \emph{``Sensitive groups: Avoid all physical activity outdoors. Move activities indoors or reschedule to when air quality will be better. People with asthma: Follow your asthma action plan and keep quick-relief medicine handy. Everyone else: Avoid long or intense outdoor exertion. Schedule outdoor activities in the morning when ozone is lower. Consider moving activities indoors.''} In short, simply giving one of six categories for a given pollutant, which are ranked on an ordinal scale, is both meaningful and useful. Note that there is a difference in how to use the air pollution scores depending on the person using them, e.g., a person with asthma as opposed to a healthy, young adult. Again, the recommendations seem legitimate for the same reason that recommendations using weight-based index $A$ and \emph{Pindex} were.

As noted, the AQI reports the overall air quality class (from green to maroon) as that of the pollutant that has the highest AQI. Basing overall air quality only on the pollutant with the highest AQI can lead to problems. Consider whether we should adopt a new policy that is expected to produce AQI scores for the five pollutants of interest $PM$, $CO$, $SO_2$, $NO_2$, and
$O_3$ of (25,25,301,25,25) or one that is expected to produce AQI scores of (300,300,300,300,300). The worst score of the first puts this in the ``Hazardous'' (maroon) category, with other pollutants being in the ``good'' category. The worst in the second puts this at the high point of the ``Very Unhealthy'' (purple) category for all pollutants. Isn't this much worse than the first example, especially since the difference between readings of 300 and 301 might be the result of small measurement errors? In that case, to say that the air is worse in the first case than in the second might be meaningful, but useless. The value of the information is minimal, since it doesn't provide evidence for an action.  But it is presumably legitimate. There are many similar questions one could ask. For example is (25,25,325,25,25) really worse than (25,25,301,301,25), when the latter has two pollutants in the hazardous region?

Comparing the two vectors (25,25,301,25,25) and (300,300,300,300,300) suggests that averaging the AQI scores of the five pollutants and reporting that as the overall air pollution value might be more useful. Suppose we call this value the \emph{BQI}. Then to say that BQI today is higher than it was yesterday is meaningful because, analogously to our earlier discussion, changing volume measurement from cubic meter to cubic feet or cubic meter to cubic kilometer would not result in changes of the mapping between concentration and AQI values, whether at the boundaries or in between via interpolation. However, consider a much simpler air pollution measure that values each pollutant $i$ on a Likert scale using values 1, 2, 3, 4, or 5 for the value $L(i)$ for $i$, and where we report the arithmetic mean of the values over all pollutants as an overall measure of air quality $L$, i.e.,

\[L = 1/5\sum_{i=1}^{5}L(i)\]

\noindent This Likert scale of 1 to 5 is likely just an ordinal scale, so, as noted earlier, comparison of arithmetic means is meaningless. However, comparison of average (arithmetic mean) values might be more useful than comparison of highest values. It would likely be legitimate, though we observe below that it might not be in certain countries at different levels of development.

From a purely measurement-theoretic point of view, since volume is measured on a ratio scale, to say that the overall AQI today is 20\% higher than it was yesterday, or twice as high as yesterday, is meaningful. But is this useful? If AQI was 50 yesterday and it doubles to 100 today, the air goes from good to moderate. But if AQI was 100 yesterday and it doubles to 200, the air goes from moderate to unhealthy. So, the doubling conclusion has different interpretations for different levels, and this conclusion seems to be useless. If we go back to the pollution measure $L$ above, then to say that $L$ today is twice what it was yesterday is meaningless, and it is very likely that this doubling conclusion would also be useless. It would likely be legitimate, though again it could be illegitimate in certain countries at a different level of development.

Taiwan uses the U.S. EPA version of the AQI. In a study of air pollution in the Kaoping region of Taiwan, \cite{Chengetal2004} concluded that between 1997 and 2001, the average annual AQI declined from 68.5 to 62.0. Again, this is a meaningful conclusion. It also illustrates the concept of Measurement Invariance discussed in Section 2.3.  If we change the volume scale, the AQI is unchanged, and so the average is also unchanged. So, it is meaningful to say that the average AQI in one year is less than it was in an earlier year. It is even meaningful to say that it is 20\% less. But, just as with the conclusion about doubling of AQI, or decrease of AQI by 20\% being useless, so is the conclusion that the average has decreased by 20\%. Great care needs to be taken when using index numbers to justify policy changes.

As mentioned earlier, variants of the AQI are in use around the world, sometimes involving fewer pollutants. The World Health Organization (WHO) has published air quality guidelines for Europe since 1987, using a variant of AQI. It is intended to be used worldwide, with the following proviso: \emph{``Air quality standards are an important instrument of risk management and environmental policy, and should be set by each country to protect the health of its citizens. The standards set in each country will vary according to specific approaches to balancing risks to health, technological feasibility, economic considerations and other political and social factors. This variability will depend on the country's level of development, capability in air quality management and other factors. The guidelines recommended by WHO acknowledge this heterogeneity and recognize in particular that, in formulating policy targets, governments should consider their own local circumstances carefully before using the guidelines directly as legal standards''} \citep[][]{Who2005}. This suggests that different conclusions using AQI might be legitimate in one country and not in another.
\cite{Chengetal2004} make a similar point. They observe that the AQI is used by a variety of countries, but there are \emph{``differences in standard concentrations, average times, calculations, and statistical analysis''} between countries. This again suggests that conclusions using AQI might be legitimate in one country and not another.

\subsection{Ambiguity and Eclipsicity}\label{aqi2}

In a series of papers beginning with \cite{Ott1978}, authors have studied ways to minimize \emph{ambiguity} of conclusions from air pollution indices, situations when an index reports air to be highly polluted when it is not, and to minimize \emph{eclipsicity} of such conclusions, situations when highly polluted air is reported as less so. The former of course raises unnecessary alarms and the latter provides a false sense of security \citep[][]{PlaiaRuggieri2011}, potentially rendering conclusions from air pollution measurement useless. Developing indices for level of ambiguity and eclipsicity would provide a way to determine the degree to which indices of air pollution are useful, and would also help in determining ways to minimize ambiguity and/or eclipsicity. Developing such indices remains a research challenge.

Consider again the vector of AQI scores for the five pollutants $PM$, $CO$, $SO_2$, $NO_2$, and
$O_3$ and compare the two cases (100,100,100,100,100) and (10,10,10,10,100). Both cases would give an overall AQI of 100, but the air in the latter case is surely much better since it is only ozone that is at that level and all the other pollutants have very low levels.
\cite{Chengetal2004} propose a correction to AQI producing a revised index RAQI. Among other things, the correction multiplies the AQI as measured by the maximum of the AQIs of the five pollutants by a factor involving the average value of the AQI of the five pollutants and by a second factor involving the Shannon entropy. Averaging the five AQI values would definitely distinguish between the two cases of (100,100,100,100,100) and (10,10,10,10,100). Using the Shannon entropy is intended to reduce the overall AQI score when there is a varying distribution of AQI values over the different pollutants. The Shannon entropy, invented for information theoretic applications, quantifies (in expected value) the information contained in a message (in this case a message about air pollution levels) in units such as bits. A fair coin has entropy of one bit. If a coin is unfair and you are asked to bet, you will have less uncertainty. The Shannon index $H$ of a vector $x = \langle x_1, x_2, \cdots x_n\rangle$ is defined as

\[H(x) = -\sum a_i\ln a_i \]

\noindent where

\[a_i =\frac{x_i}{\sum x_i}\]

\noindent $H(x)$ maximized if each $x_i$ is the same, but if there is a wide variation in values of the $x_i$, then the Shannon index will be smaller. The details of how RAQI is measured are beyond the scope of this paper. However, we note that $H(x)$ does not change if the unit of measurement of volume changes, since $x_i$ (AQI of pollutant $i$) does not change and therefore $a_i$ does not either.

\subsection{The World's Most Polluted Countries}

IQAir's World Air Quality report\footnote{IQAir. World Air Quality Report: Region and City $PM_2.5$ Ranking. \url{https://www.iqair.com/world-air-quality-ranking}\label{iqair}. Accessed 30/11/2024.} ranks the world's countries as to air quality. Several specific decisions underlie these rankings. They are based on the World Health Organization \citep[][]{Who2005} standard for particulate matter pollution, specifically $PM_{2.5}$. $PM_{2.5}$ is chosen because it is more prevalent than the other major pollutants in the air and because it has such a wide range of health effects. WHO sets standards for daily exposure and for annual exposure, one being associated with short-term health effects and the other with long-term health effects. Because of the great importance of the latter, it is the annual exposure standard of $5\mu/m^3$ that IQAir uses.

Average $PM_{2.5}$ readings over each hour are recorded at various locations in a given city and these are then combined to give an average annual reading for the city. For a country, these city averages are combined. However, IQAir does not treat each city equally. Rather, it gives more weight to the average $PM_{2.5}$ reading in a larger city than a smaller one. The reason given for this is as follows: \emph{``IQAir aims to present an overview of the global state of air quality in a way that is conducive for meaningful comparisons of ambient air quality conditions in different locations with an emphasis on airborne pollutant exposure and the effects on human health. Consequently, a simple average calculation of all city-level $PM_{2.5}$ concentrations within an area would fail to offer meaningful insight into the relative air quality experienced by individuals across the area.''} Note that here the term ``meaningful'' would seem to correspond to our concept of ``useful,'' as opposed to the more formal concept of meaningful we have been using. Thus, while it is meaningful in our sense to say that country $A$ has a higher average $\mu/m^3$  reading than country $B$, this might not be as useful as comparing population-weighted averages of the two countries.

Specifically, if $P(i)$ is the population of city $i$ and $M(i)$ is the annual average reading of $PM_{2.5}$ concentrations for city $i$, then IQAir for country $A$ gives the metric

\[I(A)=\frac{\sum_{i\in A}M(i)\times P(i)}{\sum_{i\in A}P(i)}\]

\noindent It is easy to see that the comparison $I(A) > I(B)$ is meaningful and it seems to be useful as well. (In 2022, the country with the highest $I(A)$ was Chad, with an $I(A)$ of 89.7, which is hugely greater than the goal of 5. In 2022, according to IQAir, only 13 of the 131 countries (or regions) where data was available met the $5 \mu/m^3$  goal. For example, Australia came in at 4.2 and Iceland at 3.4.)  Note that the usefulness of the index $I(A)$ might be diminished since not all areas in a city or country have equal coverage or even any coverage by air pollution monitors, and in countries with only sparse air pollution monitor coverage such as in parts of Africa, one can wonder about the legitimacy of the comparisons using $I(A)$. (Out of over 30,000 air monitoring stations whose data was used by IQAir, only 156 were in Africa. There was only one in Chad, the country with the worst I(A).)

The conclusion that $I(A)$ has decreased by 20\% between year $t$ and year $t+1$ is also meaningful. This is because the only admissible transformation of population counts is the identity, and the admissible transformations of $M(i)$ involve multiplication by a positive constant $\alpha_i$ and all the $\alpha_i$ would be the same for consistency.

Note that IQAir gives ``live'' AQI readings at cities around the world\footnote{see footnote \ref{iqair}}, i.e. AQI readings at a given moment of time. However, it does not provide live population-weighted AQI readings for countries. If $N(i)$ is the current AQI reading for $PM_{2.5}$ for city $i$, then, as we noted with the Taiwan example, saying that $N(i)$ has decreased by 20\% from one time to another is meaningful but useless. This would also make useless the statement that the population-weighted AQI for $PM_{2.5}$ for country $A$ decreased by 20\% from one time to another: The special case where all $N(i)$ are equal and all $P(i)$ are equal reduces to saying that the common $N(i)$ has decreased by 20\% from one time period to another. Averaging $N(i)$ over a year and calculating a yearly population-weighted AQI to compare one country to another would be equally useless.

\subsection{Air Stress Index}

In contrast to AQI, which mostly looks at short-term health effects, an \emph{air stress index} \citep[][]{PlaiaRuggieri2011} takes an annual perspective. One air-stress index is of course the population-weighted $I(A)$ used by IQAir. Another air stress index considers the number of times $C_i$ in a year that the concentration of a given pollutant $i$ in the air exceeds some standard for that pollutant. It compares that to a reference value $R_i$ giving the number of times a year that is permitted in some directive or guideline. Then this air stress index, which we label $ASI$, is the average of the ratio of $C_i$ to $R_i$:

\[ASI = 1/p\sum_{i=1}^{p}\frac{C_i}{R_i}\]

\noindent  where $p$ is the number of pollutants considered. Since both $C_i$ and $R_i$ are just counts, we have what is called an \emph{absolute scale} (the only admissible transformation is the identity) and so all kinds of statements involving $ASI$ are meaningful. If $ASI(t)$ is $ASI$ for year $t$, then for example it is meaningful to say that $ASI$ increased by 20\% year over year:

\[ASI(t) = 1.2\times ASI(t-1)\]

\noindent This would be a wakeup call about air pollution and would suggest that some mitigations be put into effect, so it is useful in that sense. To determine what mitigations might be needed would require drilling down and finding out which ratio $C_i/R_i$ went up significantly year over year. Note that $ASI$ is useful for policy. It is not intended to be used to provide real-time advice on short-term health effects as AQI usually is.

We might want to set the goal of reducing $C_i$, for example the number of days for which levels of a given pollutant, e.g., ozone, are in bad AQI categories, e.g., Orange or worse. How can we tell if a given policy change has achieved a given reduction? Consider the date by which the number of such days exceeds 100 for the first time. We would like this day to be later in the year. Suppose in Year $t-1$ it is June 30, and in Year $t$ it is July 19. This is a 10\% improvement from 200 days to 180 days. But, is this meaningful? In the U.S., the federal ``fiscal year'' begins October 1, not January 1. If we use the fiscal year, the improvement is from 292 days to 272 days, about 7\%. So, the 10\% improvement conclusion is meaningless, unless we specify the beginning of the year. This shows that we need to be careful to specify additional information before drawing conclusions that we can use to make or check policy. If the beginning of the year (the zero point) is specified, then the 10\% improvement conclusion is probably not only meaningful but useful as well, at least for some purposes. It shows how much progress we are making. However, would we use it to do a cost-benefit analysis of an extra investment in air pollution control? That is not clearly a reasonable use of one observation, or even of multiple observations. More work is needed to understand usefulness and legitimacy of different applications of air pollution indices.

\section{Discussion}\label{discussion}

In the following we present some observations that arise out of the two examples we have discussed.

\begin{enumerate}

 \item \emph{Meaningful statements that are obviously false might not be useful.}

 \item \emph{Statements using index numbers are not meaningful if units used are not specified unless they are the standard units everyone uses.} Thus, it is not meaningful to say that $x$ is obese, i.e., that $BMI(x)\geq 30$, without specifying units used. The threshold 30 is set to correspond to the case where $W$ is measured in $Kg$ and $H$ in meters. Similarly, concluding that the number of days it takes in a year for a given pollutant to be in AQI categories Orange or worse 100 times has gone down might not be meaningful unless the zero point of the year is specified.

 \item A metric like percentage of body fat can be \emph{useless} because it can be expensive to obtain, difficult to measure, and difficult to standardize. The latter makes it \emph{illegitimate} as well as \emph{useless}, since one person doing the measurements might come up with a different index than another person doing the measurements, even though it is \emph{meaningful} to make comparisons such as body fatness of one person is 10\% higher than body fatness of another. \emph{Meaningfulness depends on the scales used to describe the data, not on the procedure used to gather the data or the characteristics of the population the statement using metrics or indices is describing.}

\item \emph{Usefulness includes several components: ease of use (usability), to some extent Measurement Invariance, to some extent Value of Information but also appropriateness for intended use.} BMI is easy to use, but whether using it to measure adiposity is helpful is the key to appropriateness of intended use. \emph{An index can be less useful for certain populations than for others}. For example, BMI is less useful for athletes. Along the same lines, \emph{usefulness can depend on the population a statement refers to}, and it is possible that a statement that is similar might be useful for some comparisons and not for others. For example, it can be meaningful to compare BMI for two boys of different ages or for a boy and girl of the same age, but the comparison statements are not useful because what defines overweight or obesity changes with age and gender.

     Similarly, \emph{whether a decision recommended through an index number is useful may depend on the user}, e.g., whether the user is an individual physician or a government agency, or whether a person with asthma or a healthy young adult (as with AQI). Legitimacy may also become an issue if government regulations or guidelines might be a recommendation.

 \item Using an index number like BMI to suggest medical interventions or in general actions or policies, even \emph{though meaningful and useful, may be illegitimate} for some cultures that value obesity or some religions that do not believe in taking medicine. \emph{Usefulness of a conclusion involving index numbers can depend on the decision the conclusion is used for.} For example, deciding on medical treatment based on a correlation between two indices may require a high $r^2$ but deciding on what telephone to buy may only require a much lower $r^2$.

     For certain types of correlations, it is meaningful to say that two ordinal scale, interval scale or ratio scale indices (e.g., BMI and \%BF) have a correlation of c is a meaningful statement. But whether a correlation of c is high enough to be able to conclude that one of these indices, e.g., BMI, is a \emph{useful} proxy for the other, e.g., \%BF, may depend on the discipline, with \emph{higher correlations needed for usefulness in physics and medicine than in the social sciences}. Using one index (such as BMI) might be more useful than using another (such as bioelectrical impedance analysis, BIA), not because of a higher correlation with another index (such as \%BF), but \emph{because of ease of calculation}. So, there are a variety of interpretations of usefulness.

 \item \emph{A conclusion and its negation, while both meaningful, can differ in terms of usefulness}. For example, BMI might not be very useful in identifying men who are actually obese, but quite useful in identifying men who are not obese, since in one study less than half of the men identified as obese in terms of \%BF were identified as obese by BMI, but over 90\% of men not obese by \%BF were correctly identified as not obese by BMI.

     \emph{Decisions made on the basis of index numbers may be useful to recommend certain kinds of behavior and not others.} For example, if you learned that BMI had a sufficiently high correlation with Fasting Blood Glucose level (a predictor of diabetes), the correlation might be sufficiently high to recommend a change of diet but not necessarily to recommend starting on a new medication. In short, the answer to whether a given recommendation is useful may depend on the significance of the decision involved.

 \item Using a variety of sub-indices (as in the air pollution indices $A$, Pindex, and AQI) is common with index numbers. But \emph{conclusions from index numbers, though meaningful, can be useless if they disregard interactions/inter-dependencies among the factors measured by the sub-indices.}

     \emph{If ways in which an index is measured are inconsistent, then even if statements using it are meaningful, they can be useless}. For example, with Pindex, severity factors differed from study to study, air quality standards were not all laid out for the same time period, and there were differing opinions as to how to extrapolate the standards to the same time period, making use of this index useless. (There are other reasons it is useless, one being that low levels of pollution on all pollutants of interest can still lead to as high a Pindex score as one very high level on one pollutant.)

 \item \emph{Where an index is used might be relevant to its legitimacy}. For example, different conclusions using AQI might be legitimate in one country and not in another because different countries use different standard concentrations, average times, etc.

     \emph{Comparing simple averages of sub-indices, even if it leads to meaningful conclusions, might not lead to useful conclusions, and some sort of weighted average may be better}. For example, comparing air pollution values averaged over different cities in a country may be meaningful, but is likely useless unless population differences and resulting burden of health effects differences are taken into account, with something like population-weighted air pollution indices.

 \item \emph{A variety of combinations of meaningful/meaningless, useful/useless, and legitimate/illegitimate can occur:}
  \begin{itemize}
   \item Conclusions being meaningful, useful and legitimate do exist. Many of the examples given in this paper fall in this category. Using BMI to determine obesity is an example, at least for some cultures. The use of AQI to make individual decisions about reducing activity under certain air pollution conditions is another.
  \item There are conclusions that are meaningful, useful, and illegitimate. An example is use of BMI to determine obesity that is illegitimate for some cultures.
   \item There are conclusions that are meaningful, useless, and legitimate. An example is use of the weight-based air pollution index A or the variant Pindex or AQI to make policy decisions in certain cases.
   \item There are conclusions that are meaningful, useless, and illegitimate. An example involves comparisons of \%BF when the latter is measured using skinfold thickness measurements, underwater weight measurement, bioelectrical impedance, or other methods that are expensive to obtain, difficult to measure, and difficult to standardize.
   \item A meaningless conclusion can be useful and either legitimate or illegitimate, e.g., if we are comparing arithmetic mean air pollution scores when using an ordinal scale.
   \item A meaningless conclusion can also be useless and legitimate or illegitimate, e.g., if we conclude that the arithmetic mean of air pollution scores measured on an ordinal scale has doubled.
  \end{itemize}
\end{enumerate}

\section{Conclusions and Future Research}

This paper only begins the attempt to understand usefulness and legitimacy of index numbers in decision making and their relation to the already-established notion of meaningfulness. We introduced and analyzed two examples (the BMI and variants of the AQI), showing that, although meaningfulness is a necessary condition for information to be ``relevant'' for a decision process, it is far from being sufficient. Both the user's (decision maker's) acceptance (reflecting usefulness), as well as the social impact of such information (reflecting legitimacy) needs to be considered.

One clear conclusion is that it would be helpful to develop criteria or metrics for degree of usefulness. One part of the challenge in developing metrics of degree of usefulness involves finding ways to measure degree of ambiguity, or situations when an index reports a dangerous situation incorrectly (such as air being highly polluted when it is not) and to measure degree of eclipsicity,  or situations when a dangerous situation is reported as less dangerous (such as when highly polluted air is reported as less so.)

What are the different dimensions of usefulness of an index? One is ease of calculation (usability). Another is appropriateness for aiding in making the kinds of decisions that need to be made. Perhaps axioms for a usefulness measure such as described by \cite{CholvyPer19} will be useful here. Under such a multi-dimensional perspective it seems important to understand (in a typical decision support setting) what the user(s) of the index are aiming to do with that information and for which purpose. This amounts to using any among the many problem structuring methods available in the literature \citep[for an example see][]{BeltonStewart2010}. However, it is also necessary to understand what types of distinctions the decision maker would like to make. Consider for instance the Air Quality Index discussion in Sections \ref{aqi1} and \ref{aqi2}. What variations on AQI would best account for examples that suggest that the vector of pollutant AQI scores (25,25,301,25,25) should get a better air pollution score than the vector (300,300,300,300,300), that the vector (25,25,301,301,25) should get a better score than the vector (25,25,325,25,25) and that the vector (10,10,10,10,100) should get a better score than the vector (100,100,100,100,100)?

What procedures for combining sub-indices make the combined index meaningful, and/or useful, and/or legitimate? We suggest considering the axioms characterizing different preference aggregation procedures (including social choice theory and multi-attribute value theory) and mapping them to different archetypes of legitimacy requirements. As an example, non-compensatory aggregation procedures will not fit policy efficiency assessments, where we need to balance, for example, between environmental impacts and costs.


There are many more questions about meaningfulness, usefulness, and legitimacy that need to be explored. We have raised more questions than we have answered, but hope that we have given the reader something to think about.

\section*{Acknowledgements}

Fred Roberts acknowledges the support of the U.S. National Science Foundation under NSF Award 1941871. Alexis Tsouki\`as acknowledges the support of the CNRS-MITI SPLIAIDE grant.

\section*{Declaration}

All authors have no conflicts of interest or competing interests. This research did not use nor generate data.

\bibliographystyle{apalike}
\bibliography{qualityquantity411}
\end{document}